\documentclass[a4paper,fleqn]{cas-dc}

\usepackage[numbers]{natbib}

\def\tsc#1{\csdef{#1}{\textsc{\lowercase{#1}}\xspace}}
\tsc{WGM}
\tsc{QE}
\tsc{EP}
\tsc{PMS}
\tsc{BEC}
\tsc{DE}
\usepackage{amsmath}
\usepackage{amssymb}
\usepackage{caption}

\begin{document}
\captionsetup[figure]{textfont = {normalfont,large}}

\let\WriteBookmarks\relax
\shortauthors{Haoqi Ye.}

\title [mode = title]{Fast FPGA algorithm for neutron-gamma discrimination}

\author[1]{Haoqi Ye}[type=editor,
                        auid=000,bioid=1]
\fnmark[1]
\author[1]{Lian Chen}
\fnmark[2]
\author[1]{Ge Jin\corref{cor1}}
\cormark[1]

\address[1]{State Key Laboratory of Particle Detection and Electronics, University of Science and Technology of China, Hefei 230026, China}

\cortext[cor1]{Corresponding author}
\ead{goldjin@ustc.edu.cn}

\begin{abstract}
Various pulse shape discrimination methods have been used to solve the neutron-gamma discrimination problem. But most of them are limited to off-line calculation due to the computation amount and FPGA performance. In order to realize real time discriminating neutron and gamma, a new algorithm based on the traditional pulse shape discrimination methods was proposed in this paper. The new algorithm takes into account the physical properties of the pulse signal, which greatly reduces the computation and dead time without losing the precision, and can work on FPGA directly. It has a good performance in the actual experiment based on CLLB scintillation detector.
\end{abstract}


\maketitle

\section{Introduction}

  Neutron sources are always accompanied by the de-excitation gamma rays whose discrimination from neutrons is basically a complicated task\cite{r1}. To solve this problem, in the liquid scintillator and plastic scintillator detectors, which have been commonly used in neutron detection recently, the pulse shape discrimination(PSD) technique is used for n/$\gamma$ ray resolution. In such applications, PSD technique is used to separate neutron interaction events from interfering $\gamma$-ray interactions by exploiting a difference in the intensity of the slow component of the light pulses in organic scintillators that are initiated by the recoil protons and electrons. In the past few decades, various PSD techniques have been developed. The earliest was PSD circuit. With the development of digital circuits, in recent years, there have been a variety of new PSD methods based on digital circuits and computer technology. such as pattern reorganization method (PRM)\cite{r2prm},  wavelet packet transform method (WPTM) \cite{r3wptm} and discrete Fourier transform method(DFTM) \cite{r4DFTM}.
	However, due to the resource limitations of ASIC and FPGA, most of the PSD method are implemented in computers. In most of the traditional pulse-shape discrimination, the data is collected by data acquisition cards, and after being preprocessed by FPGA, packaged and uploaded to the computer, then the data is calculated offline. These algorithms have problems such as wasting FPGA performance and increasing data processing time. Therefore the new PSD method is needed which is suitable for FPGA characteristics. This PSD method needs small computation, high speed, high performance, and should work online on FPGA .

\section{Traditional PSD method}
\subsection{Charge-comparison method}
In scintillators, the optical pulses excited by charged particles have different attenuation time components. The intensity ratio of the fast and slow components is related to the mass and charge of the excited particles, that is, to the ionization density formed by charged particles in the scintillator. For example, in organic scintillator stilbene crystal, the intensity ratio of the slow component to the fast component in the light pulse excited by fast neutrons is 4.5 times that of the light pulse excited by $\gamma$ rays\cite{r5CCM}.
The fluorescence emitted by the scintillator is collected by a photomultiplier tube. When the photomultiplier tube operates in a linear range, the shape of the current pulse drawn from its anode reflects the shape of the light pulse emitted by the scintillator, i.e., the current pulse can be expressed as (1):
\begin{equation}
I\left( t \right) =I_f\left( \rho \right) e^{-\frac{t}{\tau_f}}+I_s\left( \rho \right) e^{-\frac{t}{\tau _s}}
\end{equation}
In equation, $\tau_f$,$\tau_s$,$I_f(\rho)$,$I_s(\rho)$ are the decay time of fast and slow components and the maximum value of current pulses. The total charge can be obtained by integrating the current pulse with an integral loop with a large time constant
\begin{equation}
$$
Q=\int_0^{\infty}{I\left( t \right) dt=I_f\left( \rho \right)}\tau _f+I_s\left( \rho \right) \tau _s
$$
$$
=Q_f\left( \rho \right) +Q_s\left( \rho \right) 
$$
\end{equation}

The charge pulse integration is also composed of two parts: fast and slow. Neutrons and gamma rays form different $ Q_s/Q_{total}$, so this ratio is used to identify particles. This method is called charge comparison method.
In actual calculation, the parameter PSD is defined as the ratio of the integral of the trailing edge of the pulse to the integral of the whole pulse.
To quantify the n/$\gamma$ discrimination power at a given energy threshold, a figure of merit(FOM)is used. The \emph{FOM} parameter is defined as (3):
\begin{equation}
FOM=\frac{\tau}{FWHM_n+FWHM_{\gamma}}
\end{equation}
The \emph{FOM} represents the performance of event discrimination rate.
The shortcoming of this algorithm is that it requires a lot of computation and takes up a lot of resources, which is not conducive to online identification on FPGA platform.
\begin{figure}

\centering
\includegraphics[width=0.45\textwidth]{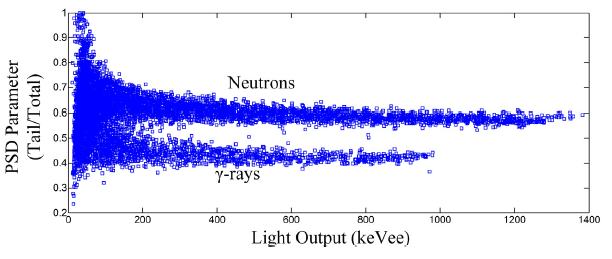}
\caption{\normalfont{The PSD performance of the system with the charge-comparison method.}}
\end{figure}
\begin{figure}
\centering
\includegraphics[width=0.45\textwidth]{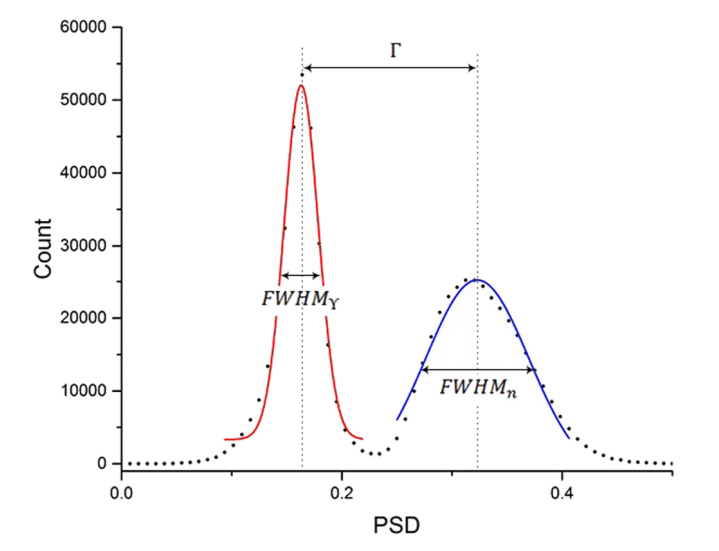}
\caption{ \normalfont{Schematic diagram of PSD spectrum.FWHM$_\gamma$ and FWHM$_n$ are half-widths of two PSD peaks.$\tau$ is the distance between two peaks.}}
\end{figure}
\begin{figure}
\centering
\includegraphics[width=0.4\textwidth]{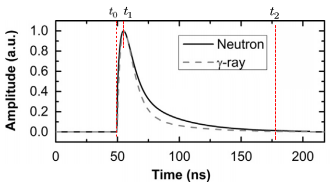}
\caption{\normalfont{The voltage signal obtained after processing the current signal of the neutron and gamma rays at the output of the photomultiplier tube. The charge comparison method uses the ratio of the integral from $t_1$ to $t_2$ of the signal waveform to the integral from $t_0$ to $t_2$ to distinguish neutrons and gamma rays}}

\end{figure}
\subsection{Pulse gradient analysis method}
Pulse gradient analysis (PGA) method is a simple discriminating method based on the different shapes of neutron and $\gamma$-ray signals\cite{r6pga}. This method selects the pulse peak and a sample point after the peak to calculate the pulse gradient. In general, the optimal value range of the time span T between the pulse wave peak and the selected sample point is 15-25ns.The specific value depends on the material of the scintillation detector and the characteristics of the photomultiplier tube.
This method has low computation and high speed, but it is very sensitive to the random fluctuation of the signal due to the great influence of noise

\section{Partial Charge-to-Peak ratio method}

In the charge-comparison method, the ratio of the back integral to the total integral of the signal was used in the identification of the fast and slow components. When calculating the PSD value, the denominator in the formula (1) plays a role of normalization. However, as the denominator is the full integral of voltage, which requires more resources, we need a substitute physical quantity.
Considering the physical meaning of the pulse waveform, the full integral at the denominator is the integral of the voltage signal generated from the operational amplifier by the pulse output from the photomultiplier tube. We can consider it as a value that is proportional to the magnitude of the charge (that is, the integral of the current waveform). From the reaction principle of neutrons and gamma rays with scintillator [5], the waveform of the normalized output current of photomultiplier tube is only related to the type of reaction. Therefore, for a definite reaction, the voltage pulse integral is the value proportional to the peak of the current waveform
In the actual calculation, $V_{peak}$, which is algebraically related to $I_{peak}$, is used. The relation between $V_{peak}$ and charge integral is (4)
\begin{equation}
\int_{t_0}^{t_2}{Vdt=\alpha V_{peak}}
\end{equation}
The definitions of $t_0$,$t_2$ are shown in Fig.2. In this condition, the peak voltage is used instead of the charge integral. This not only reduces the amount of calculation, but also accords with the physics principle during calculation.
According to the theory of fast and slow components, the fast component has a greater influence near the peak value, and the signal-to-noise ratio is poor at the tail of the pulse. So for the back integral, we don't have to take into account the integration of both ends when we integrate the pulse, which means that the equation becomes the integral from $t_1'$ to $t_2'$ as shown in Fig 4
\begin{figure}
\centering
\includegraphics[width=0.4\textwidth]{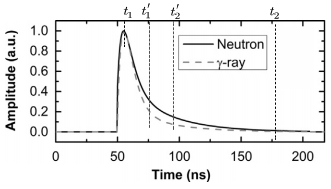}
\caption{\normalfont{The voltage signal obtained after processing the current signal of the neutron and gamma rays at the output of the photomultiplier tube}}
\end{figure}
The advantage of this modification is that for the pulse with a long trailing edge, the calculation can be greatly reduced and a lot of storage resources can be saved by selecting $t_1'$ and $t_2'$. The ratio of the back integral of the neutron/gamma waveforms before and after modification are (5) and (6): 
\begin{equation}
\eta =\frac{\int_{t_{1}}^{t_{2}}{V_{N}dt}}{\int_{t_{1}}^{t_{2}}{V_{\gamma} dt}}
\end{equation}
\begin{equation}
\eta '=\frac{\int_{t_{1}'}^{t_{2}'}{V_{N}dt}}{\int_{t_{1}'}^{t_{2}'}{V_{\gamma} dt}}
\end{equation}
Since the influence of some fast components is eliminated, the modified back-edge integral ratio is lower than the original back-edge integral ratio($\eta'>\eta$). The influence brought by noise is reduced considerably and the signal integral is basically unchanged.
Based on the physical principle of the reaction, the mathematical relationship between the physical quantities and the requirement of the calculation formula for FPGA, the PSD calculation formula of Partial charge-to-Peak ratio method on the basis of charge-comparison method and pulse gradient analysis method is shown as (7):
\begin{equation}
PSD=\frac{\int_{t_{1}'}^{t_{2}'}{Vdt}}{V_{peak}}
\end{equation}
Compared with the algorithm of the original PSD method, by setting the values of $t_1'$ and $t_2'$, 80$\%$-90$\%$ of the computation and 70$\%$ of the required storage resources can be reduced.
In order to study discrimination performance, we consider a pair of gamma rays and neutrons with the same peak current pulse. The ratios of the neutron and gamma pulse PSD values of the old and new methods are (8) and (9):
\begin{equation}
\frac{\left( PSD_{new} \right) _N}{\left( PSD_{new} \right) _{\gamma}}=\frac{\frac{\int_{t_{1}'}^{t_{2}'}{V_Ndt}}{V_{peak}}}{\frac{\int_{t_{1}'}^{t_{2}'}{V_{\gamma}dt}}{V_{peak}}}=\frac{\int_{t_{1}'}^{t_{2}'}{V_Ndt}}{\int_{t_{1}'}^{t_{2}'}{V\gamma dt}}=\eta'
\end{equation}
\begin{equation}
\frac{\left( PSD_{old} \right) _N}{\left( PSD_{old} \right) \gamma}=\frac{\frac{\int_{t_1}^{t_2}{V_Ndt}}{\int_{t_0}^{t_2}{V_Ndt}}}{\frac{\int_{t_1}^{t_2}{V\gamma dt}}{\int_{t_0}^{t_2}{V\gamma dt}}}=\frac{\frac{\int_{t_1}^{t_2}{V_Ndt}}{\alpha _NV_{peak}}}{\frac{\int_{t_1}^{t_2}{V\gamma dt}}{\alpha _{\gamma}V_{peak}}}=\frac{\alpha _{\gamma}}{\alpha_N}\eta 
\end{equation}
\par
From Fig.1, the area under $\gamma$ pulse waveform is smaller than that under neutron pulse when $V_{peak}$ is same, so $\alpha_N$>$\alpha_\gamma$ , and $\eta'>\eta$ has been known previously, and thus we can obtain (10).
\begin{equation}
\frac{\left( PSD_{new} \right) _N}{\left( PSD_{new} \right) _{\gamma}}\ >\ \frac{\left( PSD_{old} \right) _N}{\left( PSD_{old} \right) _{\gamma}}
\end{equation}
In addition, this algorithm reduces the influence of noise. Theoretically, the PSD performance of Partial charge-to-peak ratio method should be better than that of charge-comparison method. Compared with the pulse gradient method, the calculation formulas of the two methods are similar. However, considering the noise, random error and other factors, the Partial charge-to-peak ratio method is more stable than the pulse gradient method under the condition that the calculation amount is not too different.

\section{FPGA Programs}
\subsection{The optimization of Algorithm}
In FPGA, we express the waveform integral as the sum of the values of the discrete points from ADC. Because the integral position depends on the peak position, we need to find the peak before we integrate the pulse. In this case, the pulse signal should be preserved first which requires a lot of FPGA resources and a long dead time for integration. Three registers used to store the peak, the difference between current time and peak time, and the integral value are used to solve this problem. In order to reduce the dead time of the system, integration is carried out at the same time as receiving pulse signals. Setting the value of ADC input to FPGA as $V_i$, when $V_i$ exceeds the set threshold $V_T$, the program starts to work. We use a register to record the time difference between the current signal value and the peak value. When the time difference is between $T_1'$ and $T_2'$, the current signal value is added to the value in the register used to record the integral. If the current signal value exceeds the peak value, the values in register for time difference and register for integral are cleared to zero, and the peak is updated to the current value. When the value in the time register exceeds the preset pulse length $T_e$, the PSD parameter value should be calculated. The PSD parameter value is defined as integral divided by peak. The $V_T$, $T_1'$, $T_2'$, $T_e$ are defined as Fig.5 shows. 
\begin{figure}
\centering
\includegraphics[scale = 1.1]{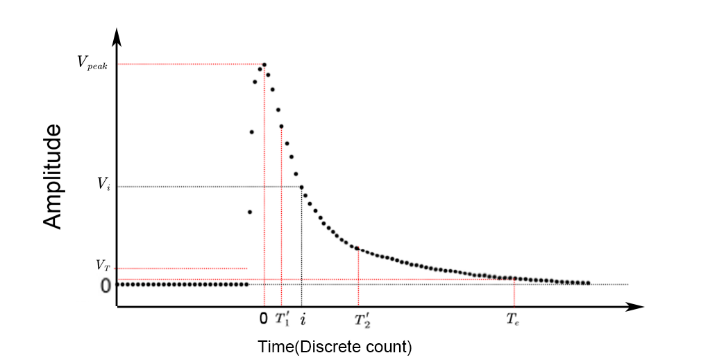}
\caption{\normalfont{Discrete pulse signal which FPGA gets from ADC. $T_1'$ and $T_2'$ are the upper and lower limits of the integration interval, and $T_e$ is the pulse length, usually set as the time at which the signal value is reduced to 5$\%$ of the peak value. i is the time difference between the current time and the peak time, $V_{peak}$  is the current peak value. }}
\end{figure}
\par
Under this optimization, it is not necessary to record the whole waveform signal. Only a few register are needed to record the time difference, signal maximum value and integral. Considering the requirements of online work, only two judgment statements are considered per clock time, and only one or two calculations are performed to minimize the dead time. This design can not only save a lot of resources, but also perfectly accord with the characteristics of small amount of computation and short dead time required by FPGA.
\subsection{FPGA state machine}
According to the algorithm, the FPGA state machine is shown in Fig6
\begin{figure}
\centering
\includegraphics[width=0.45\textwidth]{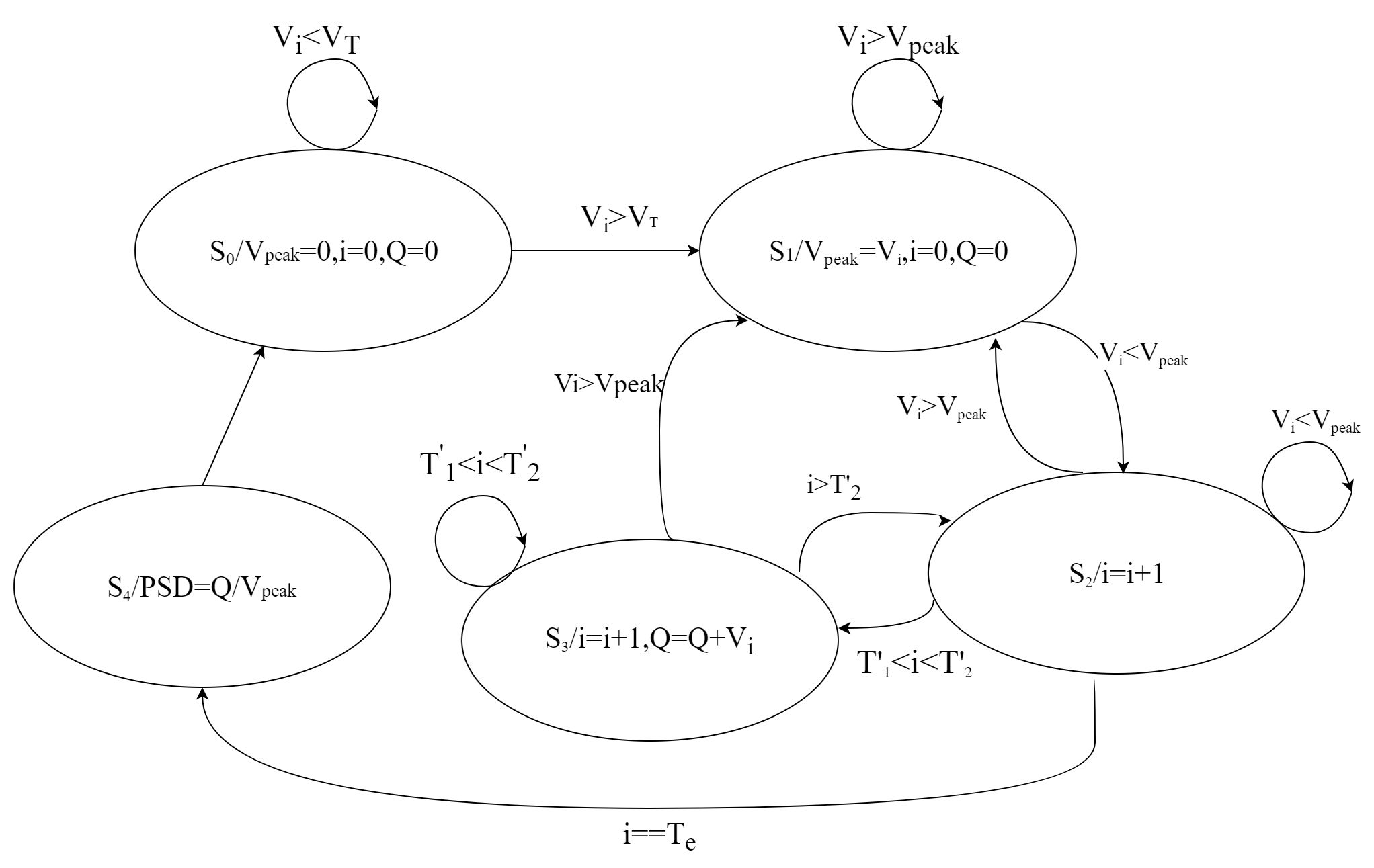}
\caption{\normalfont{FPGA state machine, $S_0$-$S_5$ are state codes, $V_i$ is the current signal value, and Q is the integral. Other parameters are the same as in Fig.5.}}
\end{figure}

\section{Experimental results and discussion}
We used the experimental platform as Fig.7 shows. This design uses a 1.5inch*1.5inch CLLB crystal packaged with a R9420 PMT made by Hamamatsu Company. The readout board reads the charge signal from photocathode, and then transfer the signal to Front End Board (FEB). After being shaped and amplified by FEB, the signal is digitized by Data Collect and Process (DCP) board.
\begin{figure*}
\centering
\includegraphics[width=0.9\textwidth]{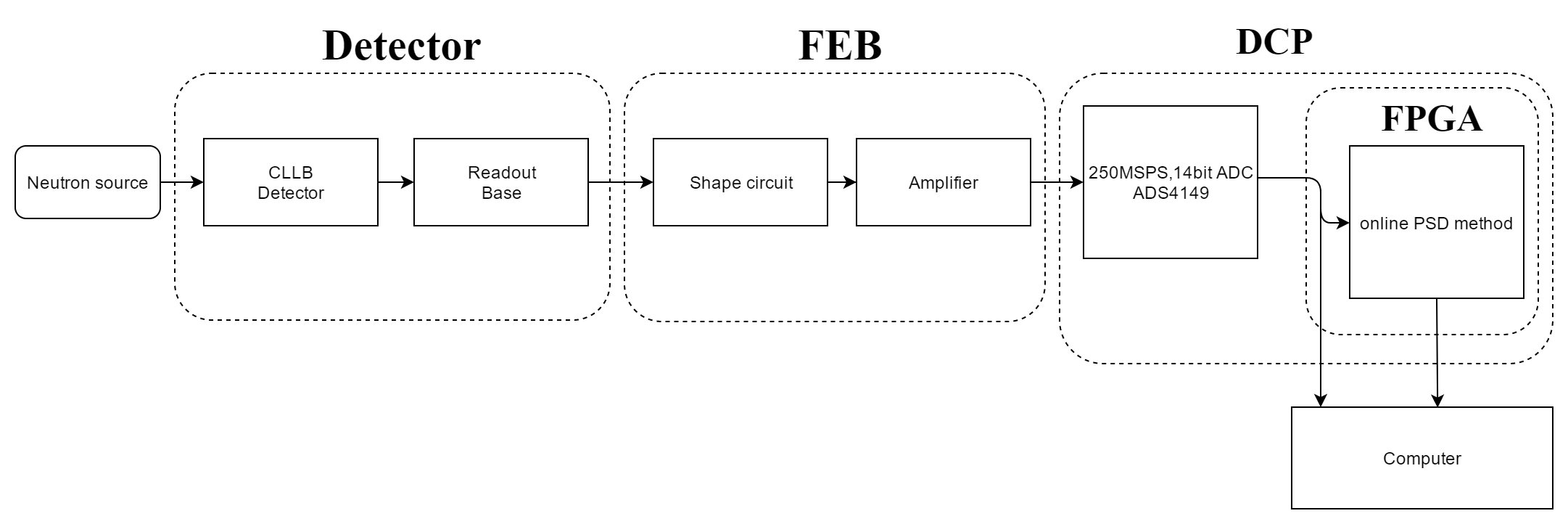}
\caption{\normalfont{Experimental platform based on CLLB crystal packaged with PMT}}
\end{figure*}
\par
A 250MSPS,14bit ADC in DCP board is used to digitize the signal pulse, then online PSD methods deployed by a FPGA process the data stream. At the same time, FPGA upload the data stream to computer. We use other offline PSD method to process the data in computer. In the FEB, in order to shape the original waveform ,we used a SK filter to spin wave to suit our 250MSPS ADC. In this case, the trailing edge of the pulse is about 1000ns(95$\%$-5$\%$)because the ADC sampling rate is not high.
To analyze the properties of different PSD methods, we used Partial Charge-to-peak ratio method and pulse gradient analysis method as online methods, used charge-comparison method as offline method.
For the $\varDelta t$ parameter of gradient analysis method, the most appropriate value 600ns, which is 150 clock in ADC, was selected after many experiments. For the $T_1',T_2'$ of partial charge-to-peak ratio method, we chose $T_1'$=150(600ns), $T_2'$=180(720ns), $\varDelta T$=30(120ns). The energy threshold was chosen the same value (300keV). Considering the balance of precision and resources, we selected 1024 channels precision in FPGA, that is, there were 1024 channels for storing PSD parameter. The value of the divided charge integral was shifted to the left before the division to accommodate the multichannel.
The PSD spectrums of these three PSD methods are shown in Fig.8-10,The \emph{FOM} parameter was determined by fitting a double Gaussian function to the distribution of the PSD parameter.

Due to the lead shielding between the detector and the neutron source in the experiment, the detector received a very small dose of gamma rays, resulting in the \emph{FOM} of the PSD energy spectrum in Gaussian fitting much smaller than that without lead shielding
As can be seen from the results in Fig.6-8, partial charge-to-peak ratio method has better performance compared with gradient analysis method and charge-comparison method. In addition, it has a small amount of computation and adapts to the characteristics of FPGA platform, which indicates that this method is of great practical significance. 
Another study of this method is to consider the values of parameters. We need to consider energy threshold, the place where start the integration $T_1'$ and the place where end the integration $T_2'$. From the noise point of view,the increase in the \emph{FOM} with increasing energy threshold is apparent. About the influence of other parameters, the larger $T_1'$ is, the smaller the effect of the fast component is; the smaller $T_2'$ is, the smaller the effect of noise is, larger the $T_2'-T_1'$ value is, more stable the performance is (the smaller the influence of random error) but also the greater the computational effort is. In order to study the influence of each parameter on the \emph{FOM}, we made a series of experiments and calculations. Results are shown in Table 1.

\begin{figure}
\centering
\includegraphics[width=0.45\textwidth]{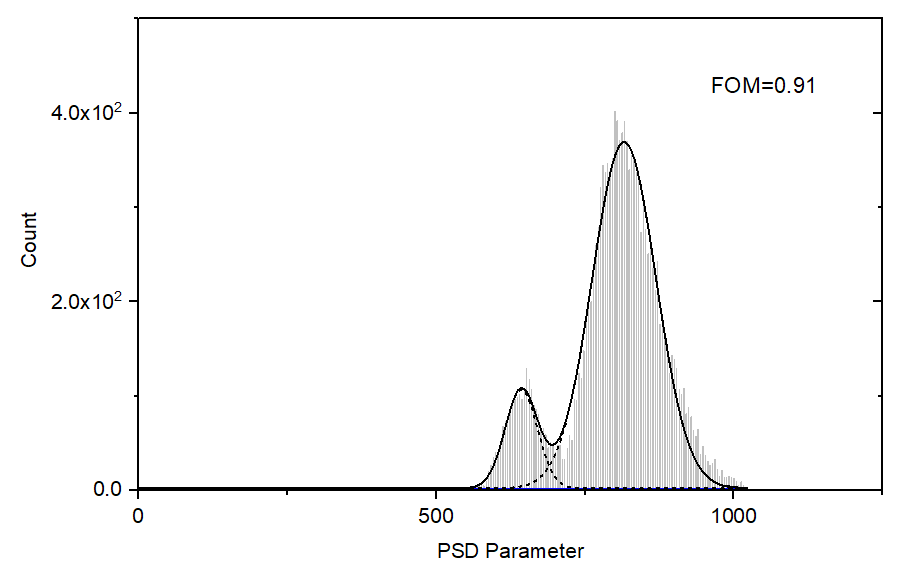}
\caption{\normalfont{Charge-comparison method's PSD spectrum with \emph{FOM}, the PSD size has been adjusted.}}
\end{figure}
\begin{figure}
\centering
\includegraphics[width=0.45\textwidth]{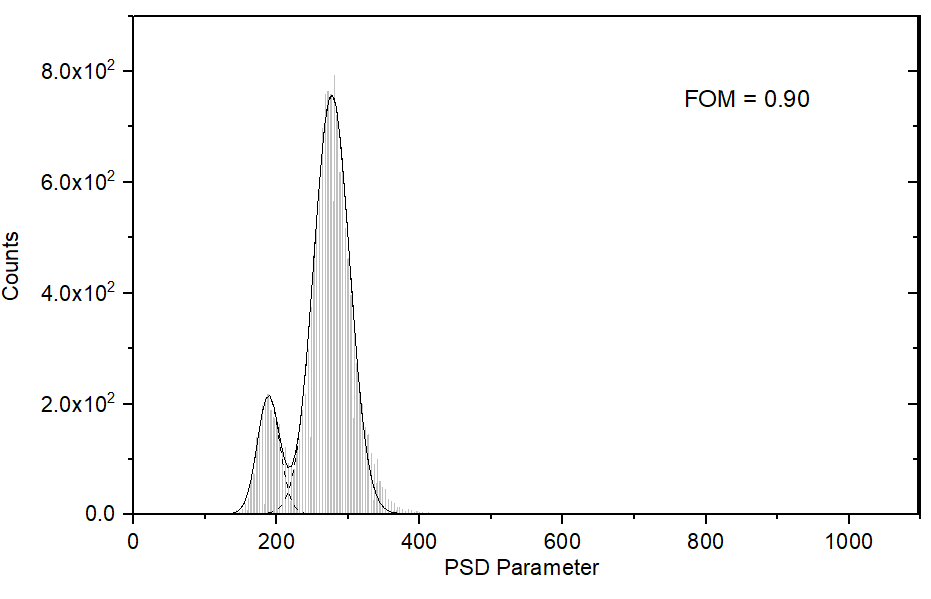}
\caption{\normalfont{Gradient analysis method's PSD spectrum with \emph{FOM}}}
\end{figure}
\begin{figure}
\centering
\includegraphics[width=0.45\textwidth]{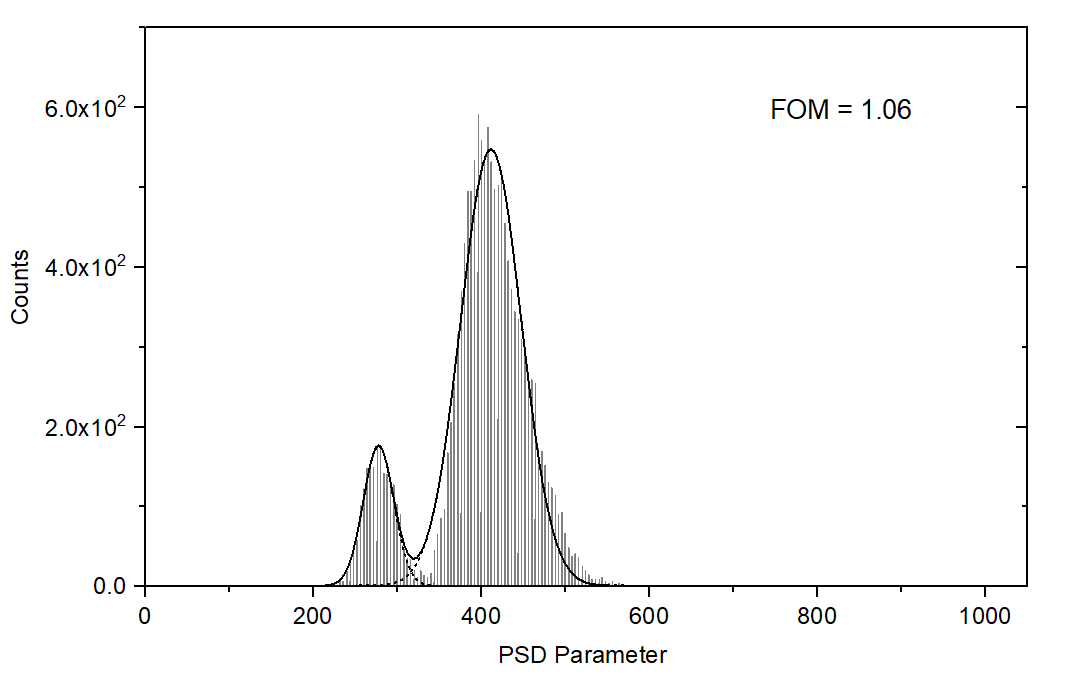}
\caption{\normalfont{Partial charge-to-current ratio method's PSD spectrum with \emph{FOM}}}
\end{figure}

\begin{table}[width=.9\linewidth,cols=4,pos=h]
\caption{\normalfont{{\emph{FOM}s obtained by selecting different $T_1',T_2'$ with threshold =300keV .The $\varDelta T$ in Table1 means $T_2'-T_1'$ .The number means how many clocks. 1 clock = 4ns ($F_{clk}$ = 250Mhz)}}}\label{tbl1}
\begin{tabular*}{\tblwidth}{@{} L|LLL@{} }
\toprule
 & $\Delta T=10$ & $\Delta T=30$ & $\Delta T=50$\\
\midrule
$T_1'=100$ & 0.75 & 0.84 & 0.91 \\
$T_1'=150$ & 1.02 & 1.06 & 1.03 \\
$T_1'=200$ & 0.9 & 1.02 & 1.00 \\
\bottomrule
\end{tabular*}
\end{table}

The selection of appropriate integral interval has a great influence on the \emph{FOM}. In this experimental platform, the appropriate integral interval is 600ns-720ns. In other experimental devices, the appropriate interval should be selected according to the detector and the forming function of the forming circuit. It is appropriate to select 10$\%$ pulse length at about 60$\%$ of the pulse falling edge (95$\%$-5$\%$).
The experimental results of the energy threshold value are shown in Table 2.

\begin{table}[width=.9\linewidth,cols=4,pos=h]
\caption{\normalfont{\emph{FOM}s under different thresholds with $T_1'$ = 150(600ns),$\varDelta T$ = 30(120ns)}}\label{tbl2}
\begin{tabular*}{\tblwidth}{@{} L|LLL@{} }
\toprule
Threshold & 300keV  & 400keV & 480keV\\
\midrule
$T_1'=150,\varDelta T=30$ & 1.06 & 1.12 & 1.16 \\
\bottomrule
\end{tabular*}
\end{table}

As we expected, the \emph{FOM} increased as the threshold increased. The selection of threshold value should be made according to the specific situation. Excessive threshold value will bring more data loss.
One limitation of FPGA is that it cannot use floating numbers. So the same algorithm has different accuracy on FPGAs and computers. When calculating the PSD value, the integral is shifted to the left first. This processing is equivalent to the precision setting of floating numbers. The algorithm was experimented on both FPGA and computer to compare the precise difference. The results in Table 3 show that there is little difference between the results obtained by computer and those obtained by FPGA.

\begin{table}[width=.9\linewidth,cols=4,pos=h]
\caption{\normalfont{\emph{FOM}s with $\varDelta T$=30(120ns), threshold = 300keV}}\label{tbl3}
\begin{tabular*}{\tblwidth}{@{} L|LLL@{} }
\toprule
           &Computer & FPGA \\
\midrule
$T_1'=100$ & 0.86 & 0.84 &\\
$T_1'=150$ & 1.07 & 1.06 &\\
$T_1'=200$ & 1.02 & 1.02 &\\
\bottomrule
\end{tabular*}
\end{table}

\section{Conclusions}
In this study, we discuss a fast n/$\gamma$ discrimination algorithm which can work on FPGA in real time with uncomplicated computation. By considering the characteristics of the fast and slow components in the pulse signal and combining the advantages of the charge comparison method and the pulse gradient method, the calculation amount were reduced 80$\%$ by this method compared with the former method. The method was tested with a CLLB crystal packaged with a R9420 PMT, and generated a good PSD spectrum online in FPGA. In the PSD resolution performance, partial charge-to-peak ratio method is better than traditional algorithms. In parameter setting, 60$\%$-70$\%$ of the falling edge is recommended as an integral interval. The algorithm gets similar results running in FPGA and computer, but has little dead time(three clocks) in FPGA, which means this algorithm is suitable for online n/$\gamma$ discrimination.

\section*{Acknowledgments}
This work was supported by the National Natural Science Foundation of China under Grant(No. 11875249).

\bibliographystyle{IEEEtran} 
\bibliography{ref}

\end{document}